\documentclass[9pt,twocolumn,twoside]{opticajnl}
\journal{opticajournal} 

\setboolean{shortarticle}{true}

\title{Machine Learning based Optimization of CV-QKD Under Practical Constraints}

\author[1,*]{Svitlana Matsenko}
\author[2]{Amirhossein Ghazisaeidi}
\author[3]{Marcin Jarzyna}
\author[3,4]{Mateusz Kucharczyk}
\author[5]{Mikkel Schmidt}
\author[3,4]{Konrad Banaszek}
\author[1]{Darko Zibar}

\affil[1,*]{DTU Electro, Technical University of Denmark, DK-2800, Kgs. Lyngby, Denmark}
\affil[2]{Nokia Bell Labs, 91300 Massy, France}
\affil[3]{Centre of New Technologies, University of Warsaw, CeNT, University of Warsaw, 02-097 Warszawa, Poland}
\affil[4]{Faculty of Physics, University of Warsaw, Pasteura 5, 02-093 Warsaw, Poland}
\affil[5]{DTU Compute, Technical University of Denmark, DK-2800, Kgs. Lyngby, Denmark}

\affil[*]{svitma@dtu.dk}

\begin{abstract}
Practical hardware limitations, including finite transmitter and receiver filter lengths as well as the finite resolution of digital-to-analog and analog-to-digital converters, lead to mode mismatch and degrade the performance of continuous-variable quantum key distribution systems. To address this, we develop a machine learning–based end-to-end optimization framework that jointly optimizes transmitter pulse shaping and receiver matched filtering. The approach employs reinforcement learning under realistic hardware constraints, including a limited number of filter taps, finite digital-to-analog and analog-to-digital converter resolution, analog low-pass filtering and the optimal mean photon number. By mitigating mode mismatch and accounting for implementation constraints, the proposed method improves overall system performance. Simulation results demonstrate enhanced secure key rates compared to conventional approaches, demonstrating the effectiveness of the proposed framework.
\end{abstract}

\setboolean{displaycopyright}{false} 

\begin{document}

\maketitle

Continuous-variable quantum key distribution (CV-QKD) has emerged as a promising approach for achieving quantum-secure communication over optical fiber networks \cite{Zhang_2024,Wang:2025sik}. Different from discrete-variable (DV-QKD), which relies on single-photon transmission, CV-QKD encodes information in the quadratures of the optical field and employs coherent detection using standard telecommunication components. 

Recent laboratory and field experiments have demonstrated stable operation of CV-QKD systems under diverse network conditions \cite{Zhang_2024,Wang:2025sik,b7833b4ec80c41d8978e04c68b0f31e3,PhysRevLett}. However, the practical realization of cost-effective CV-QKD systems requires careful optimization under realistic hardware constraints \cite{PhysRevLett,11263288}. In particular, the limited number of taps available for both the transmitter pulse-shaping filter and the receiver matched filter leads to mode mismatch, which introduces intersymbol interference (ISI) and increases excess noise, ultimately degrading the secure key rate (SKR) \cite{10925953,11263288}. Furthermore, the finite resolution of digital-to-analog converters (DACs) and analog-to-digital converters (ADCs) introduces quantization noise, which can further impair system performance.

Furthermore, a range of hardware and signal-processing constraints affects practical CV-QKD systems, in contrast to idealized theoretical models where such limitations are typically neglected. These constraints arise from non-ideal optical and electronic components, limited computational resources, finite filter lengths, and a restricted number of samples per symbol (sps). Furthermore, the achievable pulse shapes are limited by the finite bandwidth of digital-to-analog and analog-to-digital converters, as well as optical modulators. As a consequence, practical hardware does not generate perfectly smooth and well-defined pulses, but instead introduces amplitude ripple, phase distortion, and spectral broadening, deviating from the assumption of ideal temporal modes.

\begin{figure*}[t]
    \centering
    \includegraphics[width=\textwidth]{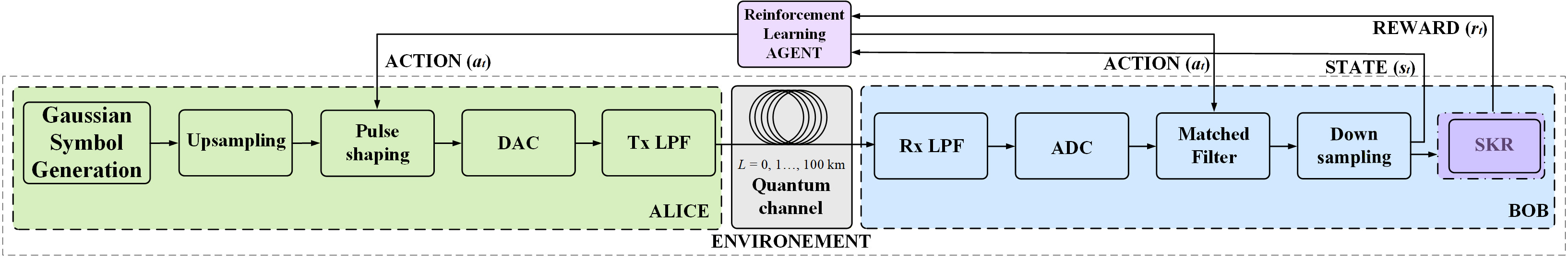}
    \caption{System architecture of the CV-QKD includes DAC – digital-to-analog converter; ADC – analog-to-digital converter; Tx/Rx LPF - low-pass-filter; SKR – secret key rate.}
    \label{fig:Fig1}
    
\end{figure*}

In this paper, reinforcement learning (RL) is utilized to jointly optimize the transmitter and receiver finite impulse response (FIR) filters and the mean photon number under practical constraints such as limited bandwidth and finite DAC/ADC resolution. For benchmarking, the proposed RL framework is compared against gradient-based backpropagation optimization. The optimizer identifies FIR coefficients that substantially mitigate overall intersymbol interference (ISI). The RL method does not require gradients of the underlying CV-QKD system model, making it more suitable for experimental implementations. Fig.~\ref{fig:Fig1} depicts the simulation setup of the considered CV-QKD system. At the transmitter, Gaussian-distributed symbols are first generated and then upsampled to four samples per symbol. The resulting sequence is processed by a finite impulse response (FIR) pulse-shaping filter with learnable weights. The filter output is subsequently converted into the analog domain using a DAC with $b$-bit resolution, followed by a fourth-order super-Gaussian low-pass filter (LPF) with a normalized 3~dB bandwidth of $0.75R$, where $R$ denotes the key rate. The generated signal is then transmitted through a fiber-optic channel with attenuation $0.2~\mathrm{dB/km}$, which is modeled as a simple loss element. At the receiver side, assuming ideal coherent detection, the signal is digitized by an ADC with $b$-bit resolution and passed through a receiver FIR filter with learnable weights. The filtered signal is then downsampled to one sample per symbol, after which the secure key rate (SKR) is evaluated based on the processed signal.

In standard CV-QKD, assuming shot-noise limited detection of both field quadratures for individual symbols and Gaussian modulation characterized by the mean photon number $\bar{n}$. Under these assumptions, the attainable SKR is given by 
\cite{Laudenbach2018,pirandola_advances}:
\begin{equation}
\mathrm{SKR} = 
\beta \log_{2} 
\left(1 + 
  \frac{\tau \bar{n}}
  {n_{\mathrm{ex}} + 1}
\right)
- g(\upsilon_{+}) - g(\upsilon_{-}) + g(\upsilon),
\label{eq:skr}
\end{equation}
where $g(x) = (x+1)\log_{2}(x+1) - x\log_{2}x$,
\begin{align}
\upsilon_{\pm} &=
\frac{1}{2} \Bigg\{
 \sqrt{
  \bigl[(1 - \tau)\bar{n} + n_{\mathrm{ex}} + 1\bigr]^{2}
  + 4\tau\bar{n}n_{\mathrm{ex}}
} \nonumber\\& \pm
\bigl[(1 - \tau)\bar{n} - n_{\mathrm{ex}}\bigr] - 1
\Bigg\},\\
\upsilon &= \frac{1 - \tau + n_{\mathrm{ex}}}
     {\bar{n}\,\tau + n_{\mathrm{ex}} + 1}
\,\bar{n},
\label{eq:nu}
\end{align}
$\beta$ is the reconciliation efficiency and $\tau$ and $n_{\mathrm{ex}}$ denote overall channel transmission and additive Gaussian excess noise, respectively. Crucially, the impact of system imperfections is captured within the latter two parameters \cite{11263288,10925953}. Note also that the channel excess noise $n_{\mathrm{ex}}$ does not include the noise contributed by the detection process, which is treated separately.

In most CV-QKD protocols the signal generated by Alice is typically assumed to be given by a train of light pulses of duration $T$
\begin{equation}
    E^{\mathrm{in}}(t) \propto \sum_{j=-\infty}^{\infty} \alpha_{j} \, u(t - jT),
    \label{eq:Ein}
\end{equation}
with amplitudes taken from the complex Gaussian distribution $\alpha_j\sim{\cal CN}(0,\bar{n})$ and orthogonal temporal waveforms $\int_{-\infty}^{\infty} dt \,  u^\ast(t - jT) u(t - j'T) = \delta_{jj'}$. In practice, however, the signal is usually generated with $s$ samples per temporal slot with the total number of $L_{\textrm{Tx}}$ samples. In such case, the mode function $u(t)$ can be written as:
\begin{equation}
    u(t) = \sum_{k= 0}^{L_{\textrm{Tx}}-1} w^{\textrm{Tx}}_k u_{\textrm{tap}}\left[t- (2k-L_{\textrm{Tx}}+1)T/(2s)\right],
\end{equation}
where $w_k^{\mathrm{Tx}}$ are weights of the individual tap functions $u_{\mathrm{tap}}(t)$ which are assumed to be real and satisfy 
\begin{equation}
    \int_{-\infty}^{\infty} dt \, u_{\textrm{tap}}(t- kT/s) u_{\textrm{tap}}(t) = \delta_{k0}
    \label{Eq:utaporth}
\end{equation}
for any integer $k$, i.e. they are temporally orthogonal. The receiver implements a FIR filter which can be similarly modeled by an effective reception mode
\begin{equation}
    v(t) = \sum_{k= 0}^{L_{\textrm{Rx}}-1} w^{\textrm{Rx}}_k u_{\textrm{tap}}\left[t- (2k-L_{\textrm{Rx}}+1)T/(2s)\right],
\end{equation}
with different set of weights $w_k^{\mathrm{Rx}}$ and total number of taps $L_{\mathrm{Rx}}$ but the same number of taps per slot $s$. 

The discrepancy between transmitter and receiver modes introduces an ISI into the QKD signal, whose impact may be characterized by a set of coefficients
\begin{equation}\label{eq:cj}
    c_j = \int_{-\infty}^{\infty}dt \, v^\ast(t-j T) u(t)=\sum_{k}w_k^{\mathrm{Tx}}w_{k-sj-(L_{\mathrm{Tx}}-L_{\mathrm{Rx}})/2}^{\mathrm{Rx}}. 
\end{equation}
The ISI affects the CV-QKD transmission in two ways. First, it introduces an additional attenuation of the symbol amplitude due to imperfect matching between the transmitter and receiver modes within a given time slot. In terms of optical power, this attenuation is described by the coefficient $|c_0|^2$ and can be incorporated into the SKR expression Eq.~(\ref{eq:skr}) by considering total effective transmision
\begin{equation}\label{eq:tau}
\tau = |c_0|^2\tau_{\mathrm{ch}},
\end{equation}
where $\tau_{\mathrm{ch}}$ is the intrinsic transmittance of the channel. The second effect of ISI is that it introduces additional noise, caused by the signal leaking from neighbouring slots, given by
\begin{equation}\label{eq:n_isi}
n_{\mathrm{ISI}}=\tau_{\mathrm{ch}}\bar{n}\sum_{j\neq 0}|c_j|^2.
\end{equation}

In realistic scenario the signal has to pass through DAC and ADC, which introduce quantization noise. Before quantization, the signal is normalized as $\tilde{x}[n] = x[n]/\max_n |x[n]|$. The DAC and ADC quantizers are modeled as $Q_b(x) = \Delta \, \mathrm{round}(x/\Delta)$, where $\Delta = 2/(2^b-1)$ is the quantization step size for a $b$-bit quantizer. The corresponding quantized signals are $y_{\mathrm{DAC}}[n] = Q_b(\tilde{x}_{\mathrm{Tx}}[n])$ and $y_{\mathrm{ADC}}[n] = Q_b(\tilde{x}_{\mathrm{Rx}}[n])$. The DAC noise is added at the transmitter and undergoes channel attenuation, yielding contribution $\tau_{\mathrm{ch}} n_d$, where $n_d = \mathbb{E}[|y_{\mathrm{DAC}} - y_{\mathrm{ref}}|^2]$ and $y_{\mathrm{ref}}$ denotes a reference signal corresponding to infinite DAC and ADC resolution. The ADC noise is introduced at the receiver with strength equal to $n_a = \mathbb{E}[|y_{\mathrm{ADC}} - y_{\mathrm{ref}}|^2]$. Both DAC and ADC contributions are modeled as additive Gaussian noise with variances $n_d$ and $n_a$ respectively. The overall excess noise at the receiver, is given by the sum of the four contributions 
\begin{equation}
n_{\mathrm{ex}} = n_{\mathrm{ch}} +n_{\mathrm{ISI}} + \tau_{\mathrm{ch}} n_d + n_a,
\label{eq:excess_noise}
\end{equation}
where $n_{\mathrm{ch}}$ denotes intrinsic channel noise.

We consider two SKR optimization procedures: the backpropagation-based SKR optimization summarized in Algorithm~\ref{alg:bp} and the REINFORCE-based optimization presented in Algorithm~\ref{alg:rl}. In both cases, the objective is to jointly optimize the FIR filters $\mathbf{w}^{\mathrm{Tx}}$, $\mathbf{w}^{\mathrm{Rx}}$, and the mean photon number $\bar{n}$ in order to maximize the SKR. For the backpropagation-based approach, gradients of the SKR with respect to the optimization parameters are computed through a differentiable system model and the parameters are updated using the Adam optimizer. The RL framework based on the REINFORCE algorithm~\cite{Williams2004SimpleSG} models the system as a policy $\pi_\theta$ with parameters $\theta={\mathbf{w}^{\mathrm{Tx}},\mathbf{w}^{\mathrm{Rx}},\bar{n}}$, while the environment captures the channel and hardware impairments. At iteration $t$, the agent updates $\theta$ and evaluates the resulting SKR. The objective is to maximize the main overlap coefficient $|c_0|^2$ while suppressing the ISI coefficients $c_j$, $j\neq0$. The state is characterized by the ISI coefficients ${c_j}$ defined in Eq.~(\ref{eq:cj}), which determine both the effective transmittance $\tau$ through Eq.~(\ref{eq:tau}) and the ISI noise contribution $n_{\mathrm{ISI}}$ through Eq.~(\ref{eq:n_isi}). The reward is defined as the resulting SKR, thereby directly linking the optimization objective to the achievable secret key rate. The policy parameters are updated according to the REINFORCE learning rule using the reward signal and a running reward baseline. 

The backpropagation-based approach relies on gradients computed through a differentiable system model, whereas the RL framework does not require differentiability and can naturally accommodate hardware impairments such as DAC/ADC quantization and other non-differentiable system components. This makes RL particularly attractive for practical CV-QKD implementations, where accurate analytical models of hardware imperfections may be unavailable or difficult to derive.

\begin{algorithm}
\caption{Backpropagation-Based SKR Optimization}
\label{alg:bp}
\begin{algorithmic}[1]
\State \textbf{Input:} Initial $w^{\mathrm{Tx}}, w^{\mathrm{Rx}}, \bar{n}$,
learning rate $\eta$, batch size $N_{\mathrm{batch}}$
\For{$b=1$ to $N_{\mathrm{symbols}}/N_{\mathrm{batch}}$}
    \State Evaluate $\mathrm{SKR}_b$ using $w^{\mathrm{Tx}}$, $w^{\mathrm{Rx}}$, and $\bar{n}$
    \State Calculate loss $L_b=-\mathrm{SKR}_b$
    \State Compute gradients $g_{\mathrm{Tx}}$, $g_{\mathrm{Rx}}$, and $g_n$
    \State Update $w^{\mathrm{Tx}}$, $w^{\mathrm{Rx}}$, and $\bar{n}$ using Adam
\EndFor
\State \textbf{Output:} Optimized $w^{\mathrm{Tx}}$, $w^{\mathrm{Rx}}$, and $\bar{n}$.
\end{algorithmic}
\end{algorithm}

\begin{algorithm}
\caption{REINFORCE-Based SKR Optimization}
\label{alg:rl}
\begin{algorithmic}[1]
\State \textbf{Input:} Initial $w^{\mathrm{Tx}}, w^{\mathrm{Rx}}, \bar{n}$,
learning rate $\eta$, baseline coefficient $\alpha$,
trace parameter $\gamma$, 
batch size $N_{\mathrm{batch}}$
\For{$b=1$ to $N_{\mathrm{symbols}}/N_{\mathrm{batch}}$}
    \State Evaluate $\mathrm{SKR}_b$
    \State Set reward $r_b$
    \State Update baseline:
           $\bar{r}_b=\gamma\bar{r}_{b-1}+(1-\gamma)r_b$
    \State Compute reinforcement factor:
           $\rho_b=r_b-\bar{r}_b-\alpha$
    \State Update $w^{\mathrm{Tx}}$, $w^{\mathrm{Rx}}$, and $\bar{n}$
           using REINFORCE
\EndFor
\State \textbf{Output:} Optimized $w^{\mathrm{Tx}}$, $w^{\mathrm{Rx}}$, and $\bar{n}$.
\end{algorithmic}
\end{algorithm}

\begin{figure}[ht]
\centering
\includegraphics[width=\linewidth]{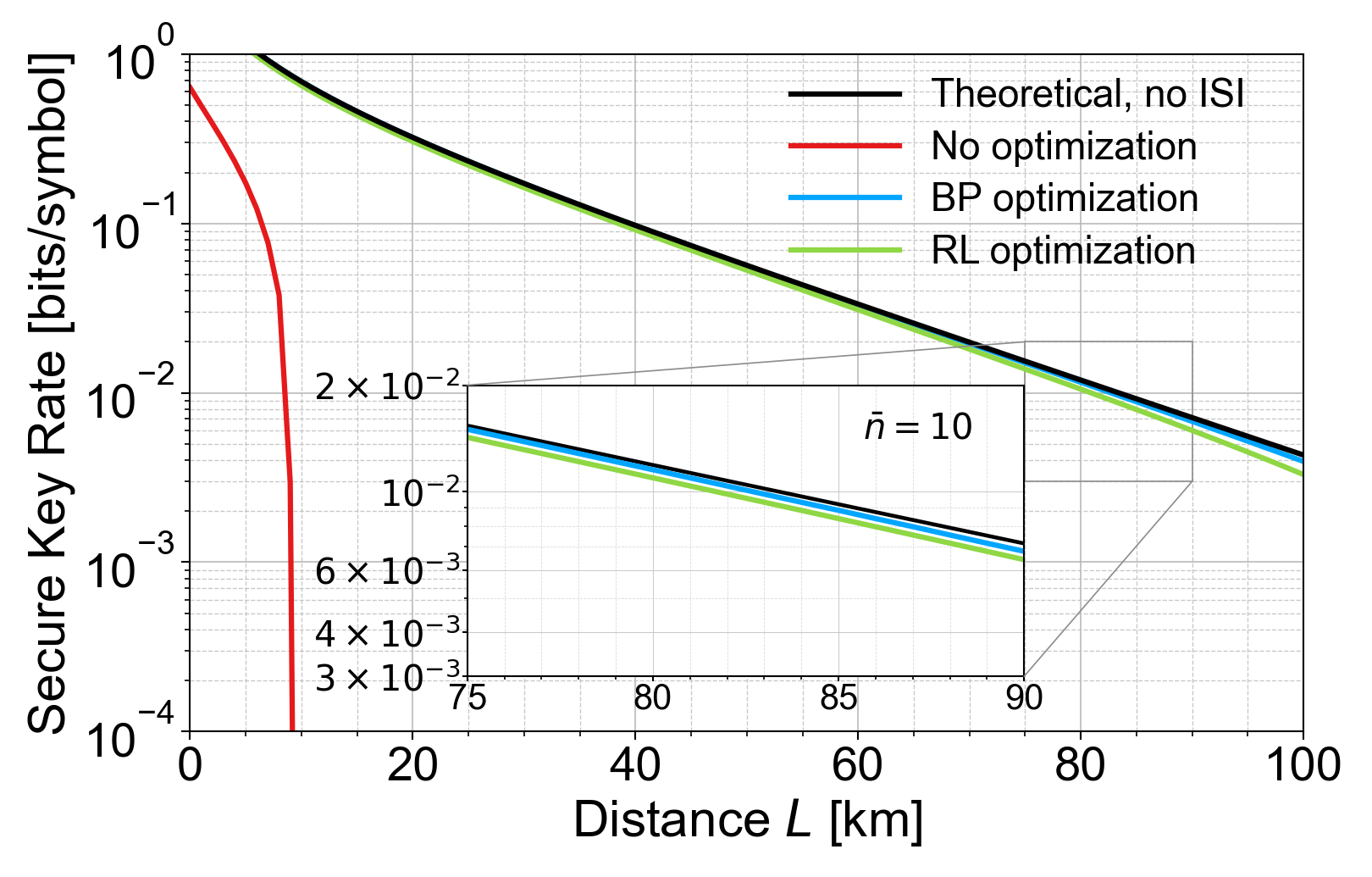}
\caption{SKR versus transmission distance with backpropagation for $\bar{n} = 6$ and $\beta = 0.95$.}
\label{fig:Fig2}
\end{figure}

Fig.~\ref{fig:Fig2} shows the SKR versus transmission distance for backpropagation and REINFORCE-based optimization with 13/101-tap Tx/Rx FIR filters, $n_{ch}=n_d=n_a=0$, and an optimized mean photon number $\bar n=6$. Backpropagation achieves a higher SKR due to the availability of gradients.

\begin{figure}[ht]
\centering
\includegraphics[width=\linewidth]{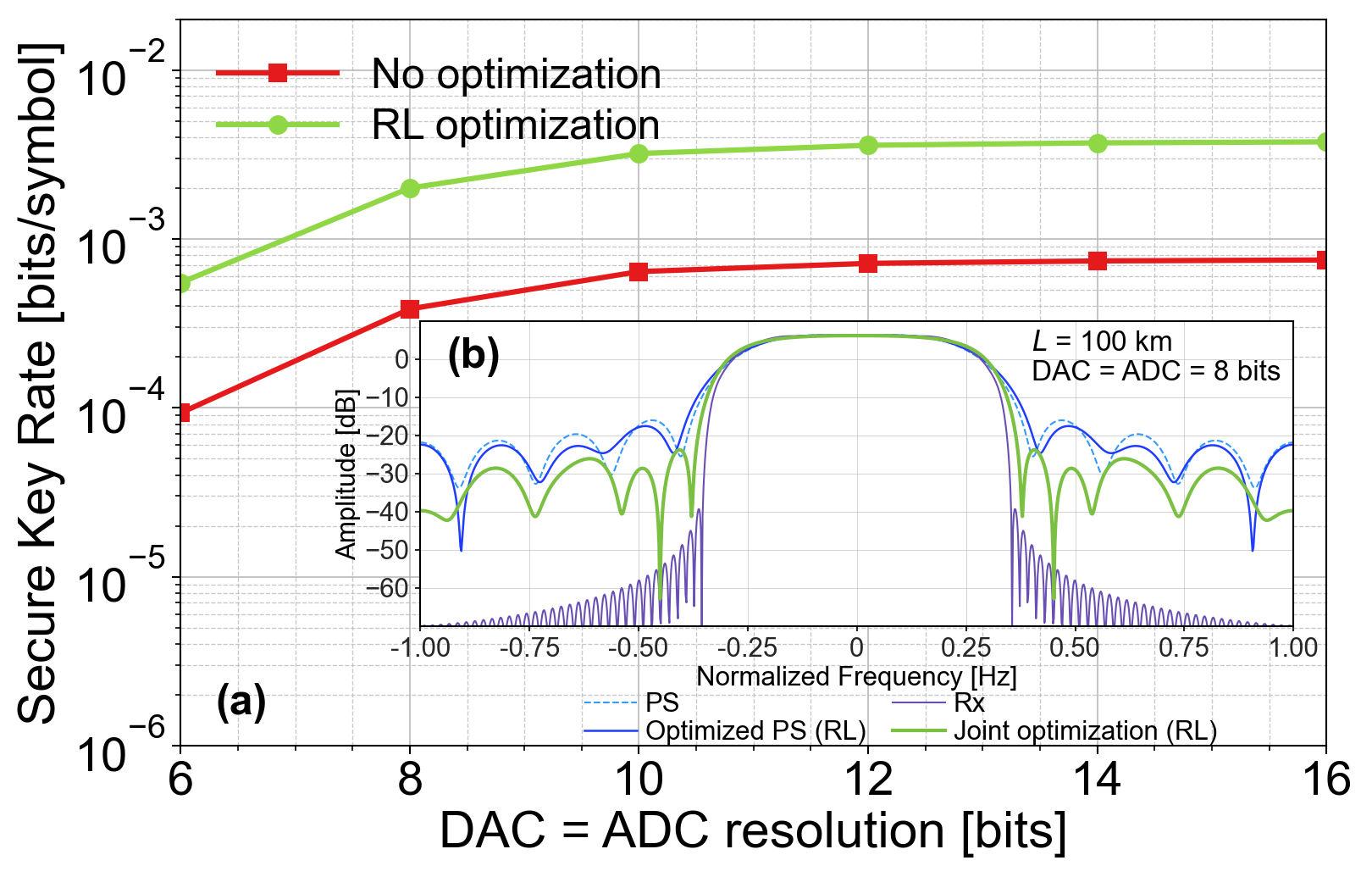}
\caption{(a) SKR vs DAC/ADC resolution; (b) Inset: filters amplitude response vs normalised frequency. }
\label{fig:Fig3}
\end{figure}

\begin{figure*}[t]
    \centering
    \includegraphics[width=\textwidth]{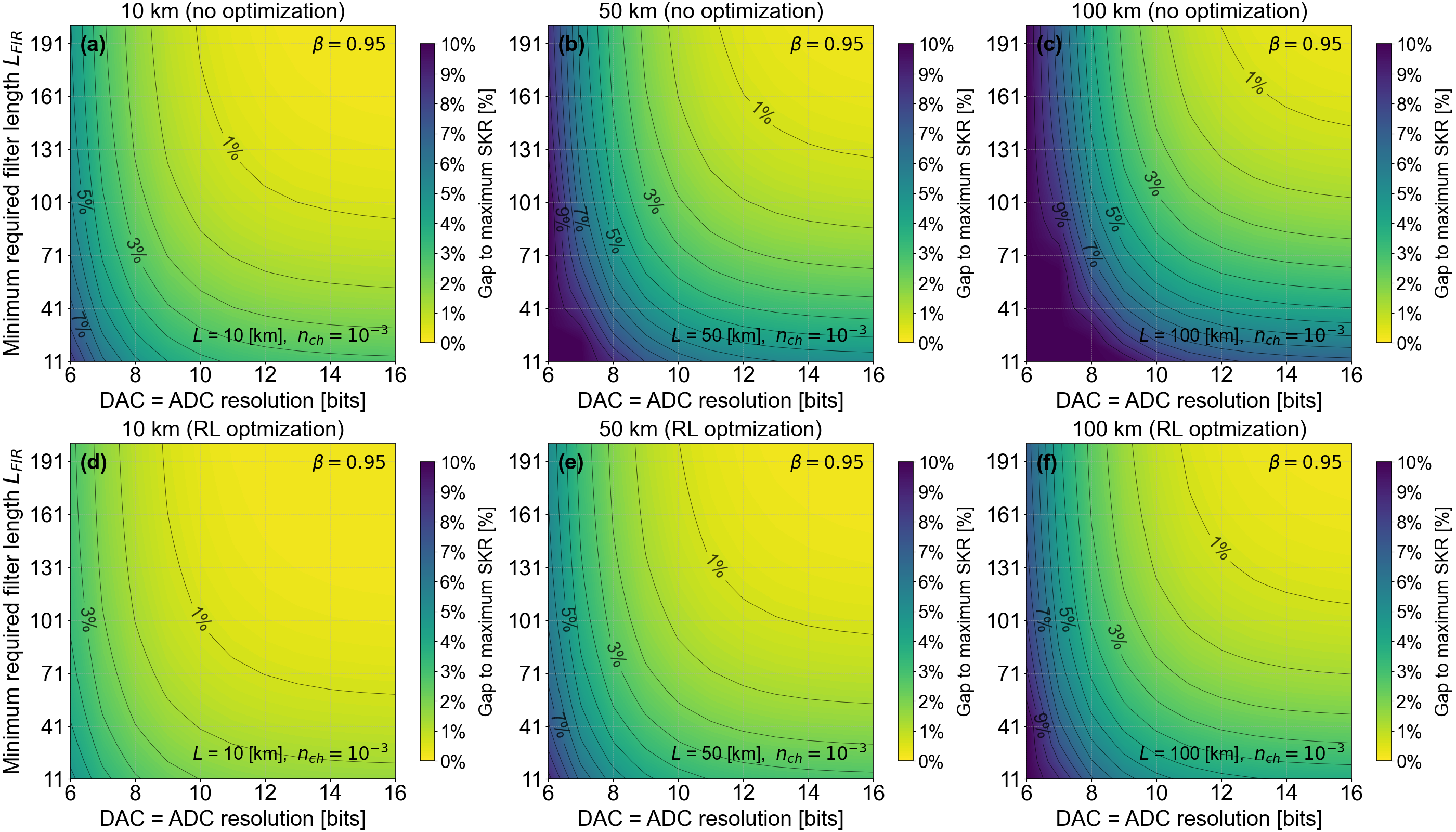}
    \caption{Relative gap to the maximum SKR as a function of the DAC and ADC resolution and the filter length of the pulse shaper and matched filters, $L_{\mathrm{PS}} = L_{\mathrm{Rx}} = L_{\mathrm{FIR}}$, for transmission distances $L \in \{10, 50, 100\}\,\mathrm{km}$, with optimized mean photon number $\bar{n} = 6$.}
    \label{fig:Fig4}
\end{figure*}

In Fig.~\ref{fig:Fig3}(a), the SKR is presented as a function of the DAC/ADC resolution. The transmitter and receiver filters have 11 and 101 taps, respectively. In the unoptimized scenario, root-raised cosine filters are used at both the transmitter and receiver. As observed in Fig.~\ref{fig:Fig3}, joint optimization of the transmitter pulse-shaping filter, receiver matched filter, and mean photon number leads to a substantial improvement in SKR compared to the unoptimized case.

In Fig.~\ref{fig:Fig4}(a)--(f), the relative gap to the maximum SKR (corresponding to infinite DAC/ADC resolution and filter length) is shown as a function of the DAC/ADC resolution (assumed equal) and the filter length of the pulse-shaping and matched filters, $w_{\mathrm{TX}} = w_{\mathrm{Rx}} = L_{\mathrm{FIR}}$, for transmission distances of $L = 10, 50,$ and $100~\mathrm{km}$. The channel excess noise is fixed at $n_{\mathrm{ch}} = 10^{-3}$, and the mean photon number is $\bar{n} = 6$. From Fig.~\ref{fig:Fig4}(a)--(c) (without optimization) and Fig.~\ref{fig:Fig4}(d)--(f) (with learning), it can be seen that, for all transmission distances, near-optimal performance (i.e., a gap below approximately $1\%$) is achieved with filter lengths of about $20$--$40$ taps and DAC/ADC resolutions of approximately $10$--$11$ bits. Increasing either the filter length or the resolution beyond this range results in only marginal improvements. Moreover, the reinforcement learning approach consistently reduces the gap to maximum SKR, especially in the low-resolution and short-filter regimes.

\begin{figure}[ht]
\centering
\includegraphics[width=\linewidth]{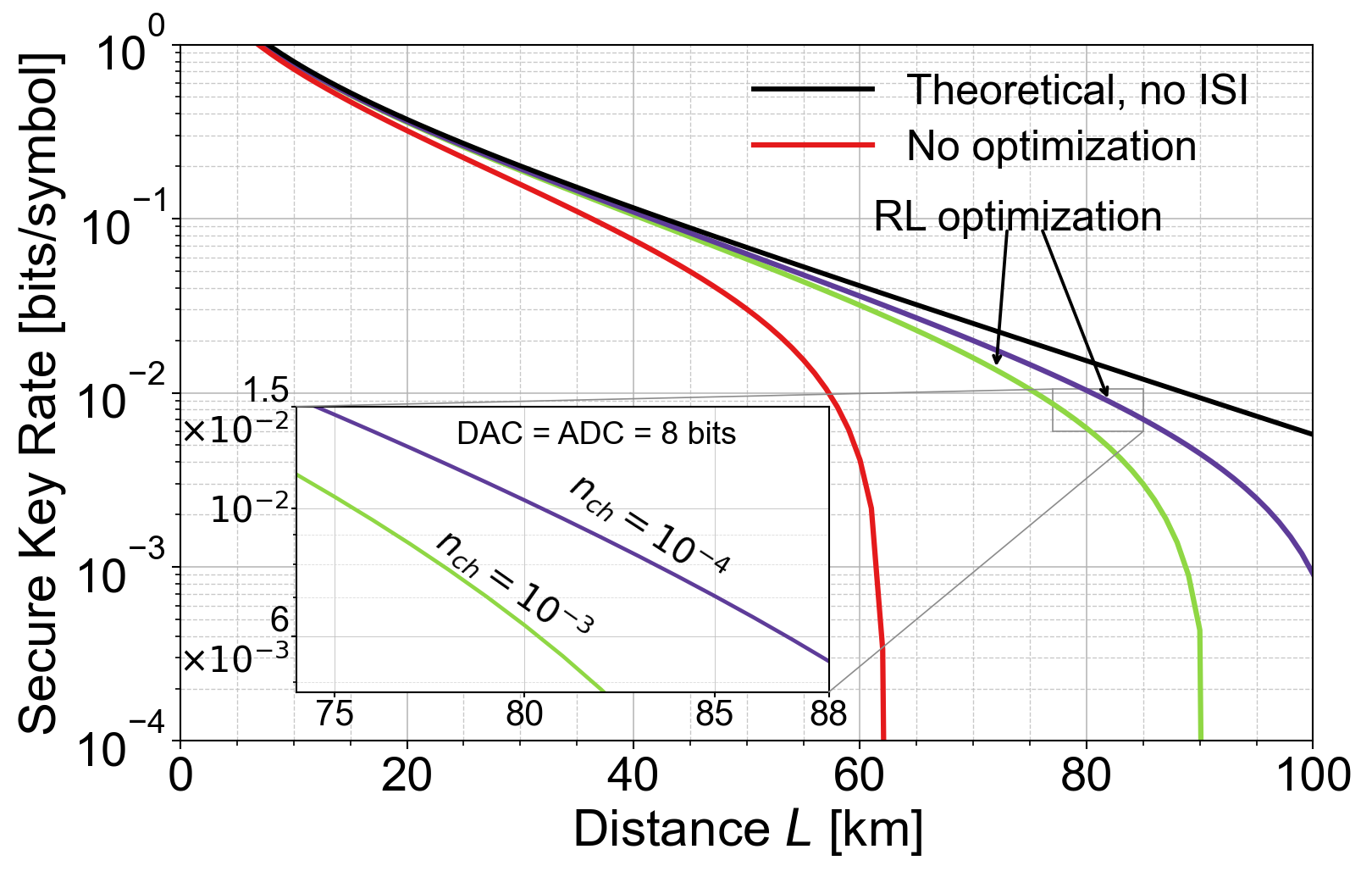}
\caption{SKR versus transmission distance for channel excess noise values $n_{\mathrm{ch}} = 10^{-3}$ and $10^{-4}$, assuming an optimized mean photon number $\bar{n} = 6$ and $\beta = 0.95$.}
\label{fig:Fig5}
\end{figure}

Fig.~\ref{fig:Fig5} presents SKR versus transmission distance for channel excess
noise values $n_{ch}=10^{-3}$ and $10^{-4}$ obtained using the
optimized mean photon number and $\beta=0.95$. The results are obtained using both analytical modeling~\cite{10925953} and RL-based optimization framework. The RL-optimized system outperforms the non-optimized case, extending the maximum transmission distance from around $60~\mathrm{km}$ to nearly $100~\mathrm{km}$ while approaching the theoretical limit over a wide range of distances. This improvement results from joint Tx/Rx filter optimization, which mitigates ISI, particularly at longer distances. The gain is particularly pronounced at longer distances, where the system is more sensitive to excess noise and filtering imperfections. The corresponding optimized mean photon number $\bar{n}$ decreases from approximately 13 at short transmission distances to about 6--8 at $100\,\mathrm{km}$, with lower values required at higher excess noise levels.

In conclusion, we demonstrate that under practical constraints, including finite filter lengths, limited DAC/ADC resolution, and finite bandwidth, joint optimization of the transmitter-receiver filters and mean photon number significantly improves the SKR. While backpropagation achieves slightly higher SKR, the RL-based approach remains competitive without requiring a differentiable model.

\begin{backmatter}

\bmsection{Funding} European Union Horizon Europe (QSNP 101114043), FENG Programme (Quantum Optical Technologies FENG.02.01-IP.05-0017/23), Foundation for Polish Science (International Research Agendas, FENG 2021–2027), Villum Fonden (VI-POPCOM VIL5448, MARBLE VIL40555).

\smallskip

\bmsection{Disclosures} The authors declare no conflicts of interest.

\bmsection{Data availability} Data underlying the results presented in this paper are not publicly available at this time but may be obtained from the authors upon reasonable request.

\bibliography{sample}



\ifthenelse{\equal{\journalref}{aop}}{%
\section*{Author Biographies}
\begingroup
\setlength\intextsep{0pt}
\begin{minipage}[t][6.3cm][t]{1.0\textwidth} 
  \begin{wrapfigure}{L}{0.25\textwidth}
    \includegraphics[width=0.25\textwidth]{john_smith.eps}
  \end{wrapfigure}
  \noindent
  {\bfseries John Smith} received his BSc (Mathematics) in 2000 from The University of Maryland. His research interests include lasers and optics.
\end{minipage}
\begin{minipage}{1.0\textwidth}
  \begin{wrapfigure}{L}{0.25\textwidth}
    \includegraphics[width=0.25\textwidth]{alice_smith.eps}
  \end{wrapfigure}
  \noindent
  {\bfseries Alice Smith} also received her BSc (Mathematics) in 2000 from The University of Maryland. Her research interests also include lasers and optics.
\end{minipage}
\endgroup
}{}

\end{backmatter}
\end{document}